\documentclass[aps,prb,showpacs,twocolumn,amsmath,amssymb]{revtex4-2}

\usepackage{amsmath}    
\usepackage{amssymb}
\usepackage{graphicx}   
\usepackage{graphics}
\usepackage{dcolumn}
\usepackage{bm}

\usepackage{multirow}

\usepackage{latexsym}
\usepackage{amssymb}
\usepackage{amsmath}

\usepackage{color}
\usepackage{epstopdf}

\usepackage[normalem]{ulem}

\begin{document}
\title{Inflation rules for a chiral pentagonal quasiperiodic tiling of stars and hexes}
\author{Viacheslav A. Chizhikov\footnote{email: chizhikov@crys.ras.ru}}
\address{NRC ``Kurchatov Institute'', A.V. Shubnikov Institute of Crystallography, Leninskiy Prospekt 59, 119333, Moscow, Russia}

\begin{abstract} 
Hexagon-boat-star (HBS) pentagonal tilings often appear in the description of decagonal quasicrystals and their periodic approximants. Being related to the Penrose tiling, they differ from the latter by a significantly higher packing density of vertices, which, in turn, depends on the relative frequency of appearance of the H, B and S tiles. Since boats (also known as ``ivy leaves'') have the lowest packing density, reducing their number in the tiling leads to an increase in its packing density. The paper proposes an inflation rule for a chiral tiling, which in principle contains no boats and therefore has the highest possible density among HBS tilings. The relationship between the tiling and the real structures of crystal approximants of decagonal quasicrystals is discussed.
\end{abstract}
\maketitle


\section{Introduction}
\label{sec:intro}

The Penrose tiling serves as an excellent example of how an abstract object can find its reflection in the surrounding world. The structure, proposed in late 1970s \cite{Gardner1977,Penrose1979}, was a response to a very interesting, if purely speculative, mathematical problem: what is the minimum set of shapes that can cover a plane only in a non-periodic manner? For almost half a century, this two-tile covering remained the best, until very recently the ultimate single-tile solution was finally found \cite{Smith2024,Smith2023}. However, the tiling did not remain merely a mathematical abstraction for long. Very soon, crystallographers noticed that it could serve as a model for ordered structures of a fundamentally new type \cite{Mackay1981}. This was a revolution that broke the paradigm that had prevailed since the 19th century, according to which the crystalline state of a substance was associated with the periodic ordering of its constituent parts. The abandonment of periodicity made it possible to expand the number of allowed point symmetries of crystals. As a result, when crystals with ``forbidden'' symmetries were actually discovered in alloys \cite{Shechtman1984}, the scientific community was theoretically prepared for their appearance \cite{Levine1984}. In contrast to ordinary periodic crystals, the new substances were called {\it quasicrystals}. To date, phases with icosahedral, octagonal, decagonal and dodecagonal symmetries have been experimentally found.

The Penrose tiling, which has plane point symmetry $10m$, as well as its generalizations can serve as a template for constructing the structures of decagonal quasicrystals \cite{Steurer2004}. The tiles simply need to be decorated with atoms or clusters of atoms. If we consider atoms (clusters) as almost spherical objects, then the problem of close-packing of spheres (disks) on a plane naturally arises. Solving this problem is important for describing and predicting the structure of quasicrystals and their periodic approximants.

Another possible application for dense pentagonal packings is in artificial 2D crystals composed of mesoscopic elements, also known as metasurfaces. Since a metasurface consists of finite size elements, it imposes a limitation on the minimum allowable distance between them. On the other hand, it is obvious that the metasurface efficiency significantly depends on the density of elements. For this reason, it is important to maximize as much as possible the vertex density of a tiling for a given minimum distance. Recently, interest in quasiperiodic metasurfaces based on the Penrose tiling has increased significantly \cite{Maslova2019,Kolmychek2020,Koreshin2022}.

We will devote Sec.~\ref{sec:history} of the article to a very brief historical overview of decagonal and pentagonal close-packing of equal disks. Section~\ref{sec:5tiling} proposes inflation rules for a pentagonal tiling of stars and hexagons (HS), which has the highest packing density among HBS tilings. Section~\ref{sec:discussion} discusses the possibility of constructing random and periodic HS-tilings.

\section{Henley, Olamy-Kl\'{e}man, and Cockayne packings}
\label{sec:history}

As is known, the most dense packing of disks of the same size on a plane is achieved on a triangular lattice with point symmetry $6m$. Its density, defined as the fraction of the total area occupied by disks, is equal to
\begin{equation}
\label{eq:maxrho6}
\rho_\mathrm{max} = \frac{\pi}{4\sin(2\pi/6)} \approx 0.9069 .
\end{equation}
In the case of decagonal symmetry, the maximum possible density (\ref{eq:maxrho6}) can be approached with a given accuracy by covering the rhombi of the Penrose tiling with small disks packed into a triangular lattice. In this case, the sides of the rhombuses play the role of linear defects, and in the limit of infinitesimal disks their contribution to the total area tends to zero. To avoid this trivial result, we impose a constraint on the allowed structures by considering only packings with decagonal local order (DLO), in which the nearest neighbors of each disk are located at the vertices of a decagon centered on that disk (all decagons are of the same size and orientation). Among structures with DLO, the periodic lattice of rhombi $T$ with an acute angle $2\pi/5$ has the highest packing density
\begin{equation}
\label{eq:maxrho5}
\rho_T = \frac{\pi}{4\sin(2\pi/5)} \approx 0.8258 ,
\end{equation}
where $\rho_T$ is the packing density of disks in rhombus $T$. However, this structure has low symmetry $2m$.

\begin{figure}[t]
	\begin{center}
		\includegraphics[width=7cm]{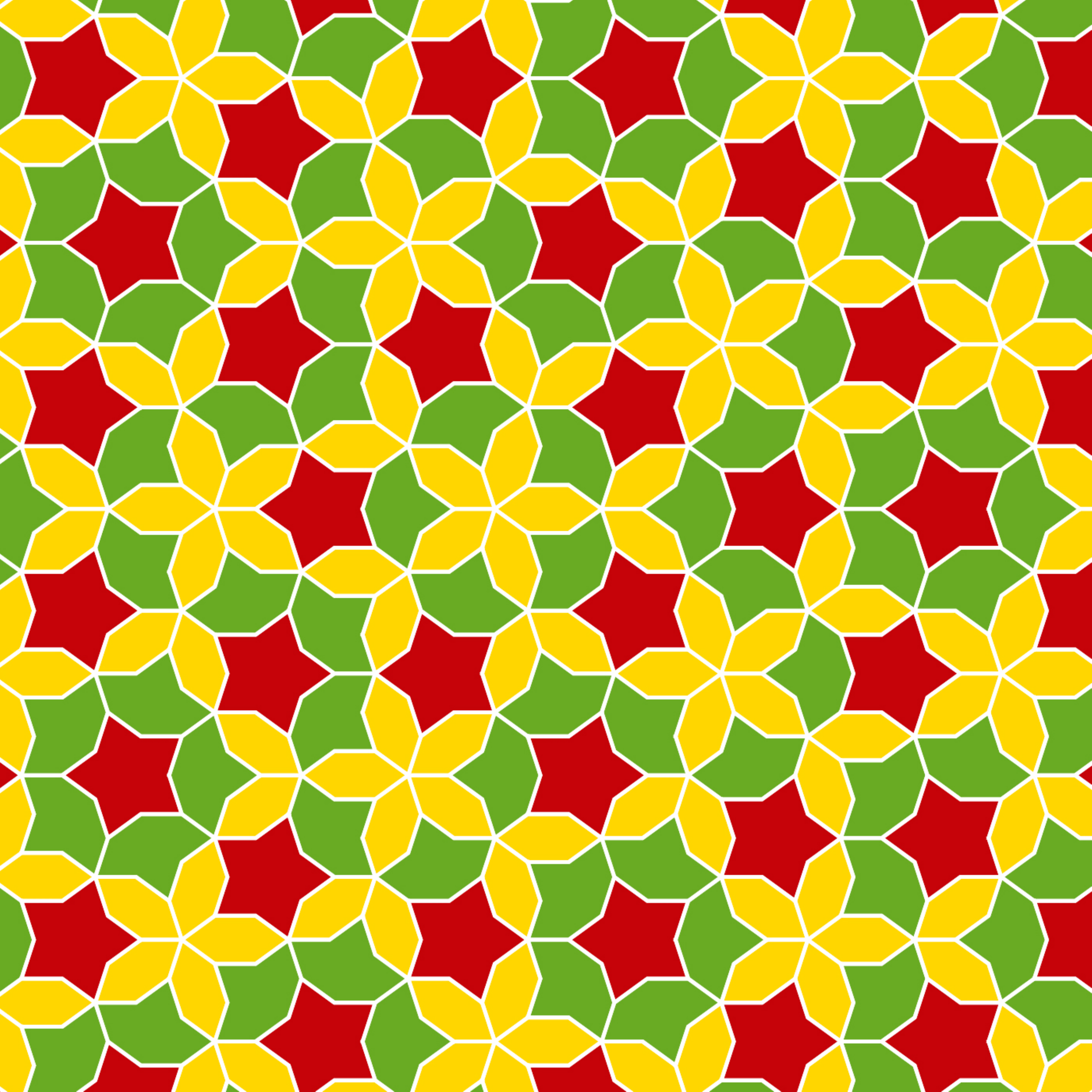}
		\caption{\label{fig:Henley} Henley tiling of ``starfish'', ``ivy leaves'' and hexes. The vertices of the pattern are those of shapes, as well as the centers of stars. The tiling is obtained from the standard Penrose rhombic pattern by eliminating some vertices lying on the short diagonals of narrow rhombi \cite{Henley1986}. Like the original Penrose pattern, the Henley tiling has point symmetry group $10m$. The frequencies of appearance of tiles are given in Table~\ref{tab:tilings}.}
	\end{center}
\end{figure}

\begin{figure}[h]
	\begin{center}
		\includegraphics[width=7cm]{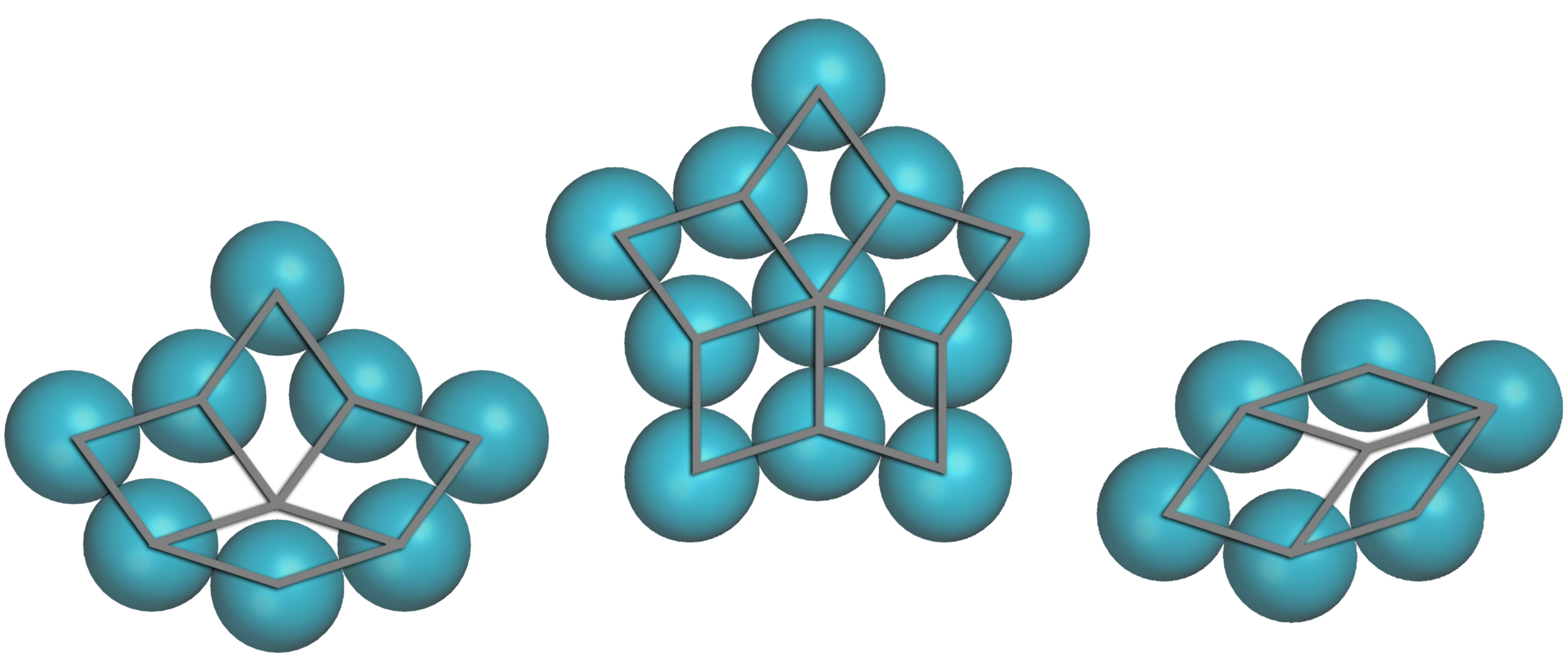}
		\caption{\label{fig:tiles} A sphere (disk) close-packing into a starfish, an ivy leaf and a hex. There are five whole disks per star, three disks per ivy leaf, two disks per hexagon. The division of the shapes into rhombuses of the Penrose tiling is shown.}
	\end{center}
\end{figure}

The first thing that comes to mind is to place the disks at the vertices of the Penrose tiling. Here, however, a slight difficulty arises: the short diagonal of a narrow rhombus $t$ with an acute angle $\pi/5$ is shorter than the edge of the rhombus, so the disks located at the ends of this diagonal overlap. In 1986, Henley proposed removing one vertex on each short diagonal \cite{Henley1986}. The result was a covering of a plane by three types of tiles, named after Lord \cite{Lord1991} ``starfish'', ``ivy leaf'' and ``hex'' (Fig.~\ref{fig:Henley}). Each shape is composed of rhombi of the original Penrose pattern: a star contains five $T$ rhombi, an ivy leaf has three $T$ and one $t$, a hexagon has one $T$ and two $t$ rhombi (Fig.~\ref{fig:tiles}). There are also alternative names for the shapes. For example, these patterns are also known as HBS (``hexagon-boat-star'') tilings \cite{Porrier2023}. The Henley tiling, like the standard Penrose tiling that generates it, has point symmetry $10m$, which is determined by a local isomorphism of the original and rotated (or reflected) tilings \cite{Levine1986,Socolar1986}. For example, Fig.~\ref{fig:Henley} shows two orientations of stars, rotated relative to each other by an angle of $\pi/5$.

Note that the term ``HBS tiling'' is sometimes used exclusively to refer to the Henley one. This article uses the term in a broader sense to refer to all coverings made from these three tiles. Various HBS tilings are used to describe quasicrystalline alloys of decagonal symmetry and their periodic approximants \cite{Cockayne1998,Cockayne2000,Steurer2001,Henley2002,Mihalkovic2002,Reichert2003,Gummelt2004,Widom2004,Mihalkovic2004,He2017}.

The packing density of disks in a tiling can be calculated as
\begin{equation}
\label{eq:rho}
\rho = \langle \rho_i \rangle = \sum_i \sigma_i \rho_i ,
\end{equation}
where the index $i$ numbers different types of tiles in the tiling, $\rho_i$ is the packing density of disks in the tile of the $i$th type, $\sigma_i = f_i s_i / (\sum_k f_k s_k)$ is the fraction of the total area occupied by the shapes of this type, $s_i$ is the tile area, and $f_i$ is the frequency of its appearance in the tiling (Table~\ref{tab:tilings}).

\begin{figure}[t]
	\begin{center}
		\includegraphics[width=7cm]{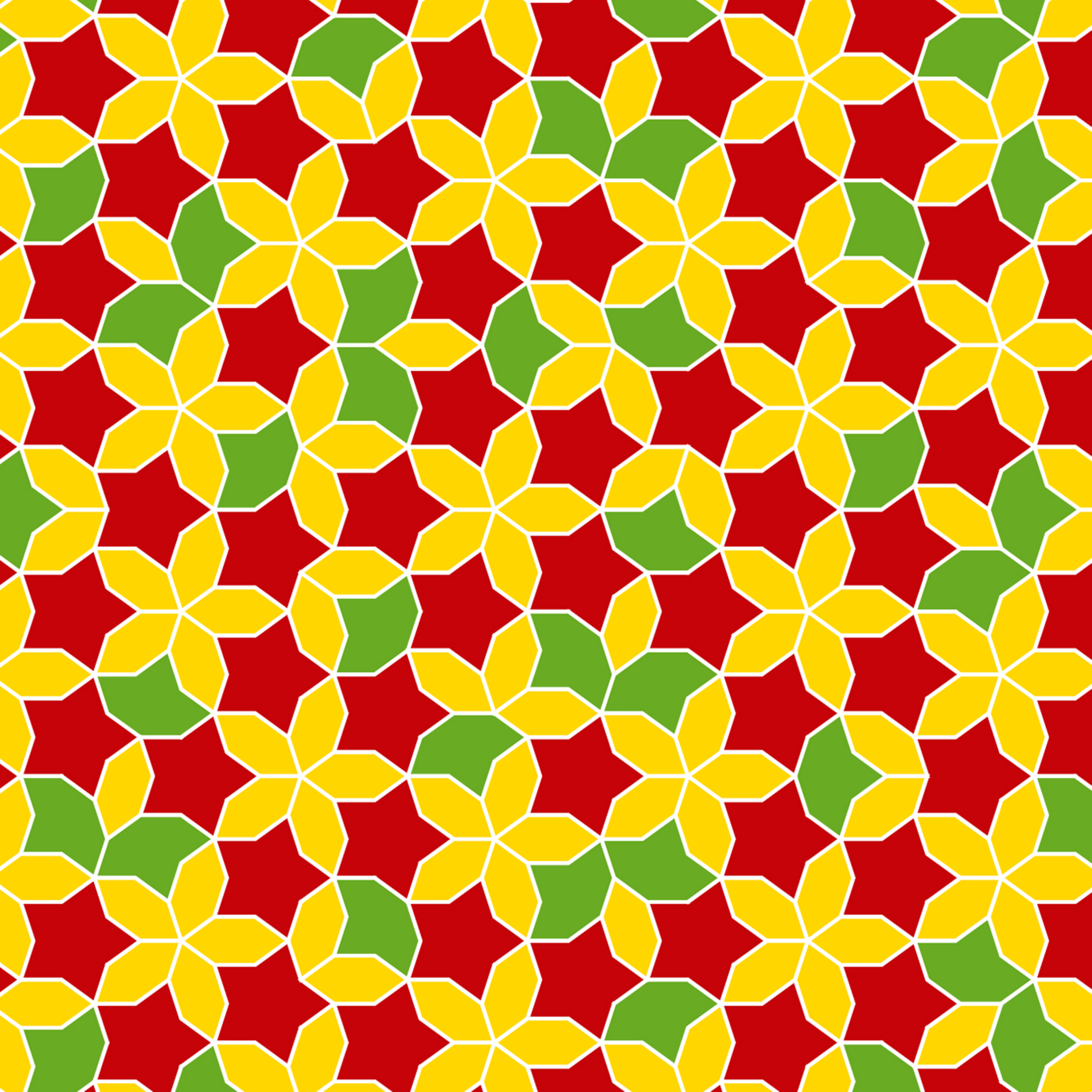}
		\caption{\label{fig:Olamy-Kleman} Olamy-Kl\'{e}man tiling of ``starfish'', ``ivy leaves'' and hexes, constructed using a specially defined acceptance domain in perpendicular space \cite{Olamy1989}. The pentagonal point symmetry $5m$ manifests itself in the same orientation of the stars. The frequencies of appearance of tiles are given in Table~\ref{tab:tilings}.}
	\end{center}
\end{figure}

\setlength\extrarowheight{5pt}
\begin{table*}
\caption{\label{tab:tilings} Area, disk packing density, and frequency of appearance of ``starfish'', ``ivy leaves'' and hexes in the Henley, Olamy-Kl\'{e}man, and Cockayne patterns. The areas of shapes and the disk packing densities are normalized to the corresponding values for the wide rhombus ($T$) of the Penrose tiling. For the Cockayne pattern, which is composed of wide rhombi and hexes, instead of a star, data are given for five rhombi $T$.}
	\begin{ruledtabular}
		\begin{tabular}{cccccc}
			\multirow{2}{10pt}{$i$} & \multirow{2}{10pt}{$s_i/s_T$} &  \multirow{2}{10pt}{$\rho_i/\rho_T$} & \multicolumn{3}{c}{$f_i$} \\
			\cline{4-6}
			&  &  & Henley \cite{Henley1986} & Olamy-Kl\'{e}man \cite{Olamy1989} & Cockayne \cite{Cockayne1994} \\
			\hline
			hex & $1 + 2\tau^{-1}$ & $2 / (1 + 2\tau^{-1}) \approx 0.8944$ & $\tau^{-1} - \tau^{-5}$ & $\tau^{-1}$ & $\tau^{-1} + \frac12 \tau^{-4}$ \\
			ivy leaf & $3 + \tau^{-1}$ & $3 / (3 + \tau^{-1}) \approx 0.8292$ & $\tau^{-4} + 2\tau^{-5}$ & $\tau^{-4}$ & 0 \\
			starfish/$5T$ & $5$ & $1$ & $\tau^{-3} - \tau^{-5}$ & $\tau^{-3}$ & $\tau^{-3} + \frac12 \tau^{-4}$
		\end{tabular}
	\end{ruledtabular}
\end{table*}
\setlength\extrarowheight{0pt}

It is convenient to measure the area of tiles in the areas of the wide rhombus $T$ of the Penrose pattern, $s_T = a^2 \sin\frac{2\pi}5$, where $a$ is the length of of the rhombus side. Then, taking into account that the narrow rhombus area is $s_t = a^2 \sin\frac{\pi}5 = \tau^{-1} s_T$ ($\tau = (\sqrt{5} + 1)/2$ is the golden ratio) and looking at Fig.~\ref{fig:tiles}, we can calculate the areas of the starfish, the ivy leaf and the hex. The packing density of disks in a shape of each type is equal to the number of disks that fit on the shape divided by its area. The frequencies of appearance of the tiles obey the conditions
\begin{equation}
	\label{eq:f-conditions}
	\left\{
	\begin{array}{l}
		f_\mathrm{hex} + f_\mathrm{ivy} + f_\mathrm{star} = 1 , \\
		f_\mathrm{hex} - f_\mathrm{star} = \tau^{-2} 
	\end{array}
	\right.
\end{equation}
(obviously, all $f_i$ are non-negative). The last condition is due to the fact that the ratio of the frequencies of wide and narrow rhombi in the Penrose tiling is $\tau$.

The system~(\ref{eq:f-conditions}) has more unknowns than equations, so it does not have a unique solution. Since among all the shapes the ivy leaf has the lowest disk packing density, it is possible to increase the packing density of the tiling by decreasing the frequency $f_\mathrm{ivy}$. Indeed, in 1989, Olamy and Kl\'{e}man succeeded in constructing a tiling with a higher packing density in this way (Fig.~\ref{fig:Olamy-Kleman}). To do this, they used a high-dimensional formalism, proposing a quasiperiodic tiling with a larger acceptance domain in perpendicular space than that of the Henley tiling \cite{Olamy1989}. Unlike the Henley tiling, the Olamy-Kl\'{e}man one has pentagonal point symmetry $5m$. In particular, this is manifested in the fact that all starfish have the same orientation.

In 1994, using daedal inflation rules, Cockayne constructed a decagonal tiling without ivy leaves, thereby pushing the disk packing density to its upper limit \cite{Cockayne1994}. In fact, the Cockayne pattern consists of wide rhombi $T$ and hexes, rather than starfish and hexes. However, for ease of comparison, Table~\ref{tab:tilings} shows the frequencies of appearence for the hex and a set of five rhombi ($5T$). The acceptance domain of the tiling contains unconnected regions and is a fractal with ``pinwheel'' motifs \cite{Cockayne1994}.

Using the data in Table~\ref{tab:tilings} and Eq.~(\ref{eq:rho}), it is possible to calculate the packing densities of equal disks in the Henley, Olamy-Kl\'{e}man and Cockayne tilings, respectively:
\begin{equation}
\label{eq:rho-tilings}
\begin{array}{l}
\rho_\mathrm{H} = \left[ 1 - \frac15 (\tau^{-2} + \tau^{-4}) \right] \rho_T \approx 0.7386 , \\
\rho_\mathrm{OK} = \left[ 1 - \frac15 \tau^{-2} \right] \rho_T \approx 0.7627 , \\
\rho_\mathrm{C} = \left[ 1 - \frac1{10} (\tau^{-2} + \tau^{-4}) \right] \rho_T \approx 0.7822 . \\
\end{array}
\end{equation}

Finally, we note that HBS tilings correspond to flat packings of spheres (disks). The three-dimensional quasicrystalline structure is formed by a stack of identical planes, with the spheres being replaced by atomic clusters. By the way, there is another way of densely packing spheres into a decagonal structure, in which the centers of the spheres extend beyond the plane. In this case, the projections of neighboring spheres onto the plane can intersect \cite{Wills1990,Cockayne1999}.

\section{Pentagonal tiling of starfish and hexes}
\label{sec:5tiling}

\subsection{Inflation rules}
\label{subsec:inflation}

Some deterministic quasiperiodic tilings, such as the standart Penrose one, can be constructed using the inflation method, which allows to obtain the entire pattern from an arbitrary finite-size piece of it or even from a single tile. At each step of the iterative process, every shape of the initial structure is divided into similar smaller ones. Then, each of the new small shapes should be increased (inflated) to the size of the initial tiles. The division of the original tile is carried out in accordance with its in-plane orientation. For example, in the Penrose decagonal tiling, there are ten possible orientations for each {\it arrowed} rhombus \cite{deBruijn1981}.

\begin{figure}[h]
	\begin{center}
		\includegraphics[width=5cm]{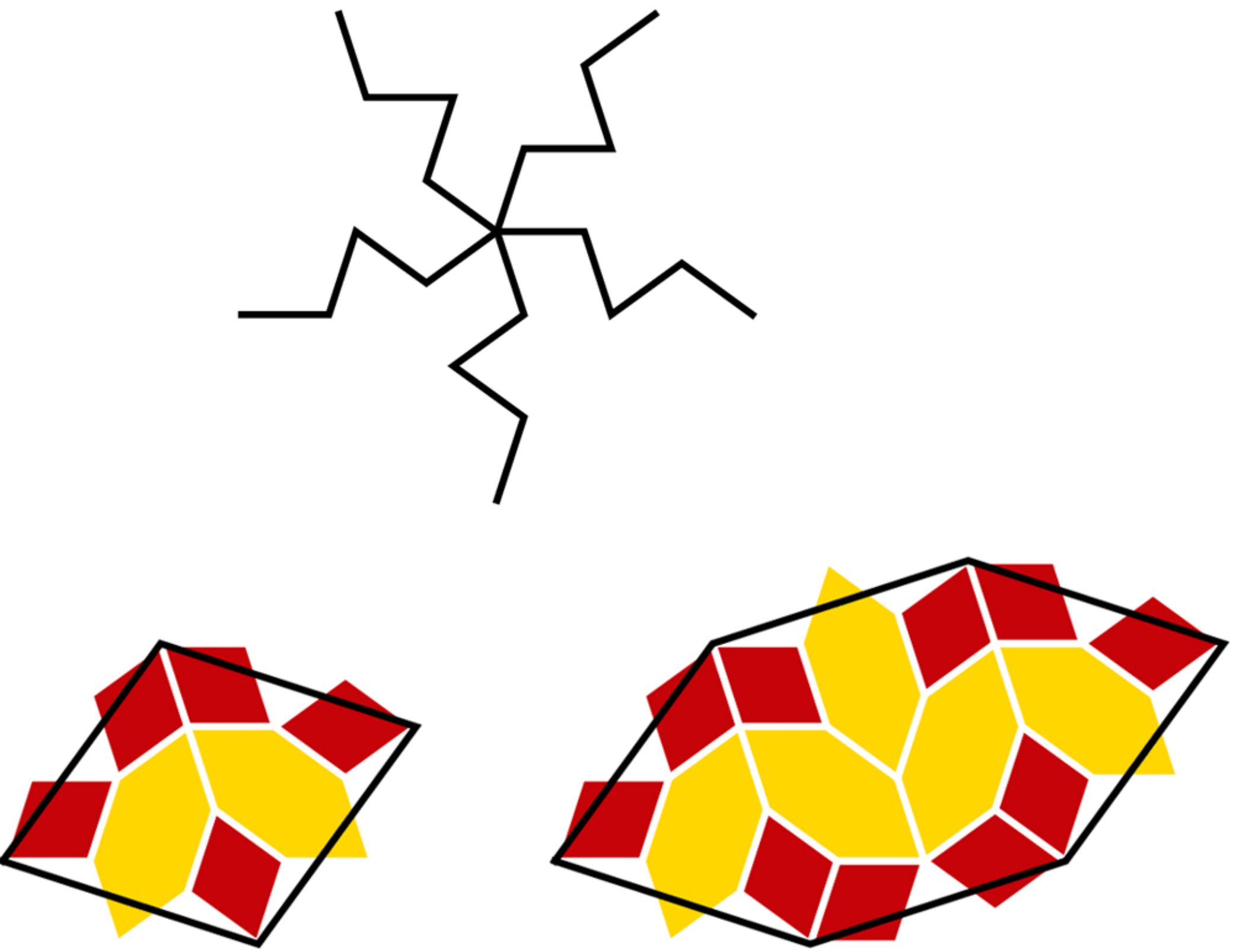}
		\caption{\label{fig:inflation} Inflation rules for rhombus $T$ and hexagon leading to emergence of a hex-starfish tiling. Inflation is carried out in accordance with the orientation of the shapes. Only one orientation of tiles out of five possible is shown; inflation rules for other orientations can be obtained using $2\pi/5$-rotations. The linear scale factor is $\tau\sqrt{2+\tau}$. Above are the matching rules of the shapes, depending only on the in-plane orientation of their common side.}
	\end{center}
\end{figure}

Figure~\ref{fig:inflation} shows the inflation rules for a wide rhombus $T$ and a hexagon, resulting in a pattern of hexes and starfish. Unlike the inflation rules of the Penrose tiling, there are only five possible orientations for each shape. The partitions for four orientations not shown in the figure can be obtained by $2\pi/5$-rotations. It is easy to see that all the smaller rhombi are assembled with the acute angles at the vertices of the original shapes. Thus, starfish of the same orientation are formed at these vertices, and the space between them is filled with hexes. The rules for matching tiles are very simple, symmetrical about the middle of the common side of the tiles and depend only on the side in-plane orientation (five possible orientations in total).

\begin{figure}[h]
	\begin{center}
		\includegraphics[width=7cm]{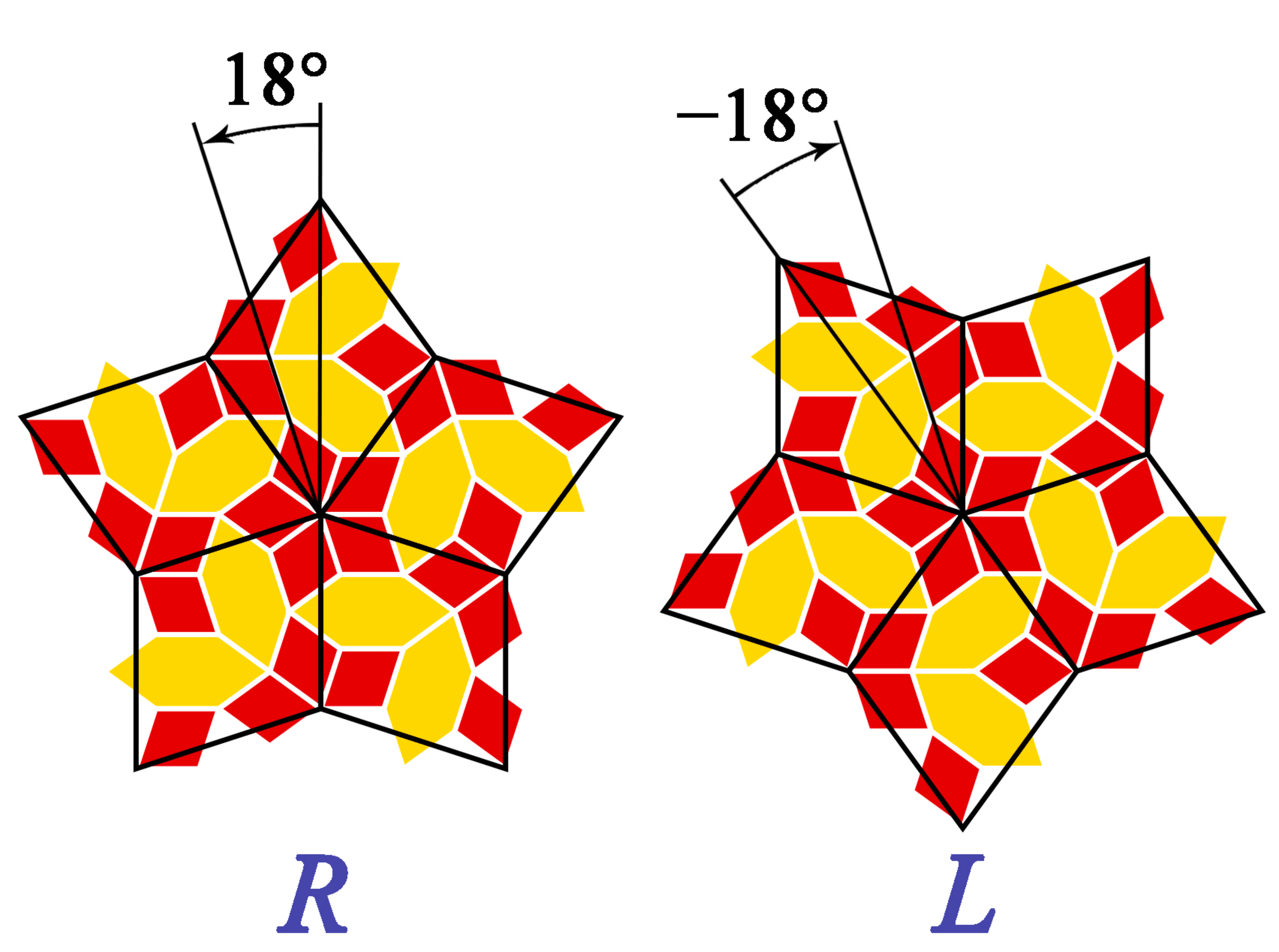}
		\caption{\label{fig:chirality} Two ways to inflate the star formed of five rhombi using the rules of Fig.~\ref{fig:inflation}. The new stars (red) are rotated relative to the original ones by an angle of $-\pi/10$ (left) or $+\pi/10$ (right), which ensures chirality of the tiling. The choice between inflation rules $L$ and $R$ can be made arbitrarily for each iteration.}
	\end{center}
\end{figure}

\begin{figure}[h]
	\begin{center}
		\includegraphics[width=7cm]{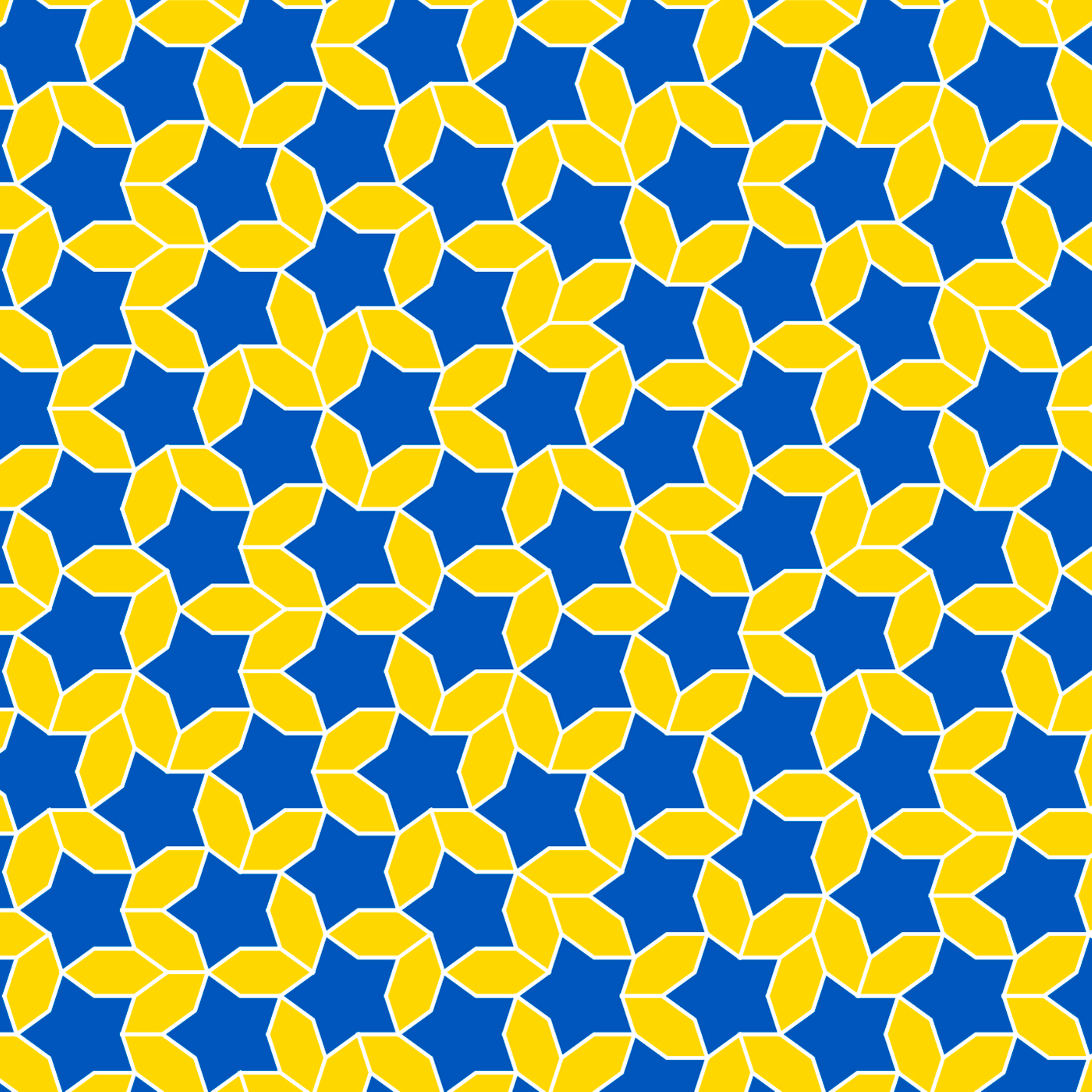}
		\caption{\label{fig:chiral-5fold} A chiral pentagonal tiling of starfish and hexes, obtained using the inflation rules from Fig.~\ref{fig:inflation}. At each inflation step the stars rotate by the angle of $+\pi/10$, which corresponds to the right boundary of the interval $[0,1]$ (see text). This tiling provides the same disk packing density as the Cockayne tiling of rhombi and hexes \cite{Cockayne1994}.}
	\end{center}
\end{figure}

There are two interesting differences between these inflation rules and those typically used for pentagonal tilings. First, the linear scale factor $s_\ell = \tau\sqrt{2+\tau} \approx 3.078$ is not a power of $\tau$. For example, for the Henley, Olamy-Kl\'{e}man and Cockayne tilings, the inflation factors are $\tau$, $\tau^2$, and $\tau^6$, respectively. Secondly, the sides of the original and inflated tilings are rotated relative to each other by an angle $\pi/2$, which is not a multiple of $\pi/5$. The last property has an important consequence. To ensure that the sides of the inflated pattern coincide in direction with the sides of the original one, before inflating, we can rotate the tiling by an angle $\pi/10$ clockwise or counter-clockwise (Fig.~\ref{fig:chirality}). It means that there is an ambiguity in the choice of inflation rules at each iteration. Thus, instead of a single tiling, you get a family of tilings. However, in appearance they will be very similar. Figure~\ref{fig:chiral-5fold} shows an example of such a tiling.

From Fig.~\ref{fig:inflation} it is easy to calculate that for a finite fragment of the tiling, the numbers of shapes of each type before and after inflation are related through a transfer matrix as
\begin{equation}
	\label{eq:transferM}
	\left( \begin{array}{l}
		n^\prime_\mathrm{star} \\
		n^\prime_\mathrm{hex}
	\end{array} \right) =
	\left( \begin{array}{cc}
		5 & 2 \\
		10 & 5
	\end{array} \right) \cdot
	\left( \begin{array}{l}
		n_\mathrm{star} \\
		n_\mathrm{hex}
	\end{array} \right) .
\end{equation}
The first of the eigenvalues of the matrix, $\lambda_1 = 4\tau + 3 = s_\ell^2 \approx 9.472$, confirms that the number of tiles grows in proportion to the area of the fragment. The corresponding eigenvector
\begin{equation}
	\label{eq:eigenvector}
	\left( \begin{array}{l}
		f_\mathrm{star} \\
		f_\mathrm{hex}
	\end{array} \right) =
	\left( \begin{array}{c}
		(\sqrt{5} - 1) / 4 \\
		\sqrt{5}(\sqrt{5} - 1) / 4
	\end{array} \right)
\end{equation}
coincides in frequencies with those of the Cockayne pattern (Table~\ref*{tab:tilings}), which means that this tiling has the same disk packing density. The ratio of the number of hexes and starfish is $f_\mathrm{hex} / f_\mathrm{star} = \sqrt{5}$.  Another eigenvalue of the transfer matrix is $\lambda_2 = \lambda_1 / \tau^6 \approx 0.5279$.

\subsection{Chirality}
\label{subsec:chirality}

As we see in Fig.~\ref{fig:chiral-5fold},all starfish in the resulting pattern have the same plane orientation. At each inflation step the new stars are rotated relative to the original ones by an angle of $\pi/10$ clockwise or counter-clockwise, depending on the choice of inflation rules (Fig.~\ref{fig:chirality}). This rotation breaks the mirror symmetry of the starfish, introducing chirality into the tiling.

Note that at each inflation we can choose either $R$ or $L$ inflation rules independently. This means that there exists a family of similar tilings with different chirality signs at different scales. This family is uncountable, and a simple correspondence can be established to the interval $[0,1]$. Depending on the history of inflation steps, each boundless tiling can be associated with an infinite sequence of letters $R$ and $L$, e.g. $RRLRLLR\ldots$, where the first letter corresponds to the last inflation step. By replacing $R$ with 1 and $L$ with 0 and adding a decimal point in front, we get the binary representation of the real number $0.1101001\ldots$, which belongs to the interval $[0,1]$. For example, Figure~\ref{fig:chiral-5fold} shows the tiling corresponding to the right boundary $1=0.(1)$ of the interval. The correspondence is not entirely one-to-one. It is violated if the string of digits ends in an infinite sequence of zeros or ones, in other words, for numbers of the form $(2k+1)/2^n$. For example, the number $5/8$ corresponds to two different tilings: $RLRLLLL\ldots$ and $RLLRRRR\ldots$, since $0.101(0)=0.100(1)$.

Two tilings are mirror images of each other if their letter sequences are related by inversion $R \leftrightarrow L$. Thus, if one of them corresponds to some $x \in [0,1]$, then the second corresponds to the number $\bar{x} \equiv 1-x$ from the same interval. For example, the reflection of the tiling $x=1$ (Fig.~\ref{fig:chiral-5fold}) is the tiling $\bar{x}=0$. We can also note an interesting mirror pair of tilings, $1/3=0.(01)$ and $2/3=0.(10)$, which transform into each other at every inflation step.

\subsection{5D formalism}
\label{subsec:5D}

To construct a quasiperiodic tiling, a multidimensional formalism is often used, where the tiling vertices are projections of nodes of a regular lattice with a dimension greater than that of physical space \cite{deBruijn1981,Kramer1984,Duneau1985,Elser1986}. For example, in decagonal local order, the nearest neighbors are separated by one of the vectors $\mathbf{b}_k = (\cos\frac{\pi k}5, \sin\frac{\pi k}5)$, $k = 0 \ldots 9$ (here and below we assume that $a = 1$). In a connected structure, one can find a chain of nearest neighbors from some initial vertex with coordinate 0 to an arbitrary one with coordinate $\mathbf{r}$. This means that the distance between vertices is determined by a set of ten integers $m_k$, $\mathbf{r} = \sum_k m_k \mathbf{b}_k$. Using the equality $\mathbf{b}_{(k+5) \% 10} = -\mathbf{b}_k$, this set is usually reduced to five integers $n_k = m_{2k} - m_{(2k+5) \% 10}$, $k = 0 \ldots 4$ (here $x \% y$ means the remainder of $x / y$). Then the coordinate of an arbitrary vertex can be written as 
\begin{equation}
\label{eq:r-parallel}
\mathbf{r} = P_\parallel \cdot \mathbf{n} = \left(
\begin{array}{ccccc}
	C_0 & C_1 & C_2 & C_3 & C_4 \\
	S_0 & S_1 & S_2 & S_3 & S_4
\end{array}
\right) \cdot \mathbf{n} ,
\end{equation}
where matrix $P_\parallel$ projects 5D vector $\mathbf{n} = (n_0, n_1, n_2, n_3, n_4)$ into physical space $E_\parallel$; $C_k \equiv \cos \frac{2 \pi k}5$, $S_k \equiv \sin \frac{2 \pi k}5$.

The 5D formalism allows one to perform operations on tilings using simple integer matrices. For example, symmetry group $10m$ of a decagon is defined by two simple generators in 5D space: reflection and rotation by an angle of $\pi/5$,
\begin{equation}
	\label{eq:m}
	m = \left(
	\begin{array}{rrrrr}
		1 & 0 & 0 & 0 & 0 \\
		0 & 0 & 0 & 0 & 1 \\
		0 & 0 & 0 & 1 & 0 \\
		0 & 0 & 1 & 0 & 0 \\
		0 & 1 & 0 & 0 & 0
	\end{array}
	\right) ,
\end{equation}
\begin{equation}
	\label{eq:R10}
	R_{10} = \left(
	\begin{array}{rrrrr}
		0 & 0 & -1 & 0 & 0 \\
		0 & 0 & 0 & -1 & 0 \\
		0 & 0 & 0 & 0 & -1 \\
		-1 & 0 & 0 & 0 & 0 \\
		0 & -1 & 0 & 0 & 0
	\end{array}
	\right) .
\end{equation}
The inflation operation defined earlier involves homothety with a coefficient of $s_\ell$ and rotation by an angle of $\mp\pi/10$. In 5D space these transformations are described by the matrices
\begin{equation}
	\label{eq:R-inflation}
	{\cal I}_R = \left(
	\begin{array}{rrrrr}
		1 & 1 & -1 & -1 & 0 \\
		0 & 1 & 1 & -1 & -1 \\
		-1 & 0 & 1 & 1 & -1 \\
		-1 & -1 & 0 & 1 & 1 \\
		1 & -1 & -1 & 0 & 1
	\end{array}
	\right)
\end{equation}
and
\begin{equation}
	\label{eq:L-inflation}
	{\cal I}_L = R_{10} \cdot {\cal I}_R = \left( {\cal I}_R \right)^T
\end{equation}
for right and left inflation, respectively.

\subsection{Acceptance domain}
\label{subsec:acceptance}

Note that the projection of all nodes of the 5D lattice is dense in physical space. Hence, the existence of a shortest distance between tiling vertices should impose some constraint on the choice of nodes accepted for projection. For the pentagonal tilings under consideration, this is expressed in the fact that the projection of selected nodes onto a subspace $E_\perp$ orthogonal to the physical one occupies a limited volume, called the acceptance domain. Let us define the projection of a node onto the space $E_\perp$ as follows:
\begin{equation}
	\label{eq:r-perp}
	\mathbf{r}_\perp = P_\perp \cdot \mathbf{n} = \left(
	\begin{array}{ccccc}
		C_0 & C_2 & C_4 & C_1 & C_3 \\
		S_0 & S_2 & S_4 & S_1 & S_3 \\
		1 & 1 & 1 & 1 & 1
	\end{array}
	\right) \cdot \mathbf{n} .
\end{equation}
It is easy to see that $z$-coordinate of the projection, $z_\perp = \sum n_k$, is always an integer, so the projection of the entire 5D lattice in perpendicular space consists of a set of equally spaced ``atomic'' planes with different $z_\perp$. Note also that the nodes of the 5D lattice, spaced apart by the vector $(1,1,1,1,1)$, are projected to the same point in physical space, so it is sufficient to use a maximum of five atomic planes, e.g. $z_\perp = 0 \ldots 4$.

\begin{figure}[h]
	\begin{center}
		\includegraphics[width=7cm]{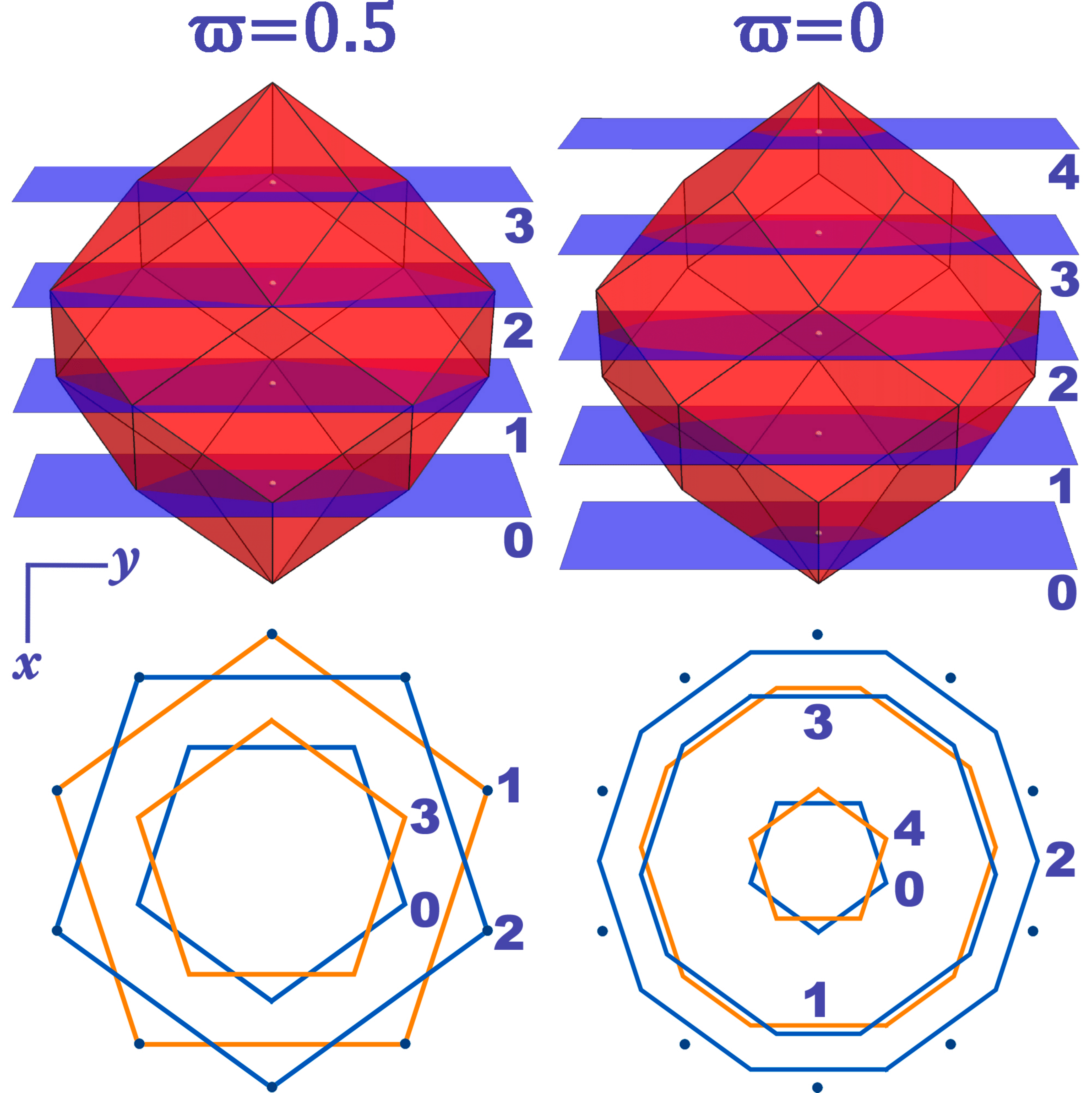}
		\caption{\label{fig:penteract} Acceptance domains of the standard (left) and a generalized (right) Penrose tilings, consisting of cross-sections of a penteract  projection by equidistant cutting planes in the perpendicular space $E_\perp$ \cite{Senechal1995}. The tilings differ in the distance $\varpi$ between the center of the penteract and the nearest plane. From the shape of the sections shown below, it is easy to see that both tilings have point symmetry group $10m$. The total cross-sectional area $\Sigma_\mathrm{P}$ does not depend on the parameter $\varpi$, see Eq.~(\ref{eq:SigmaP}). Note that all cross-sections fit inside a decagon of radius $\tau$, the vertices of which are marked with dots.}
	\end{center}
\end{figure}

\begin{figure}[h]
	\begin{center}
		\includegraphics[width=7cm]{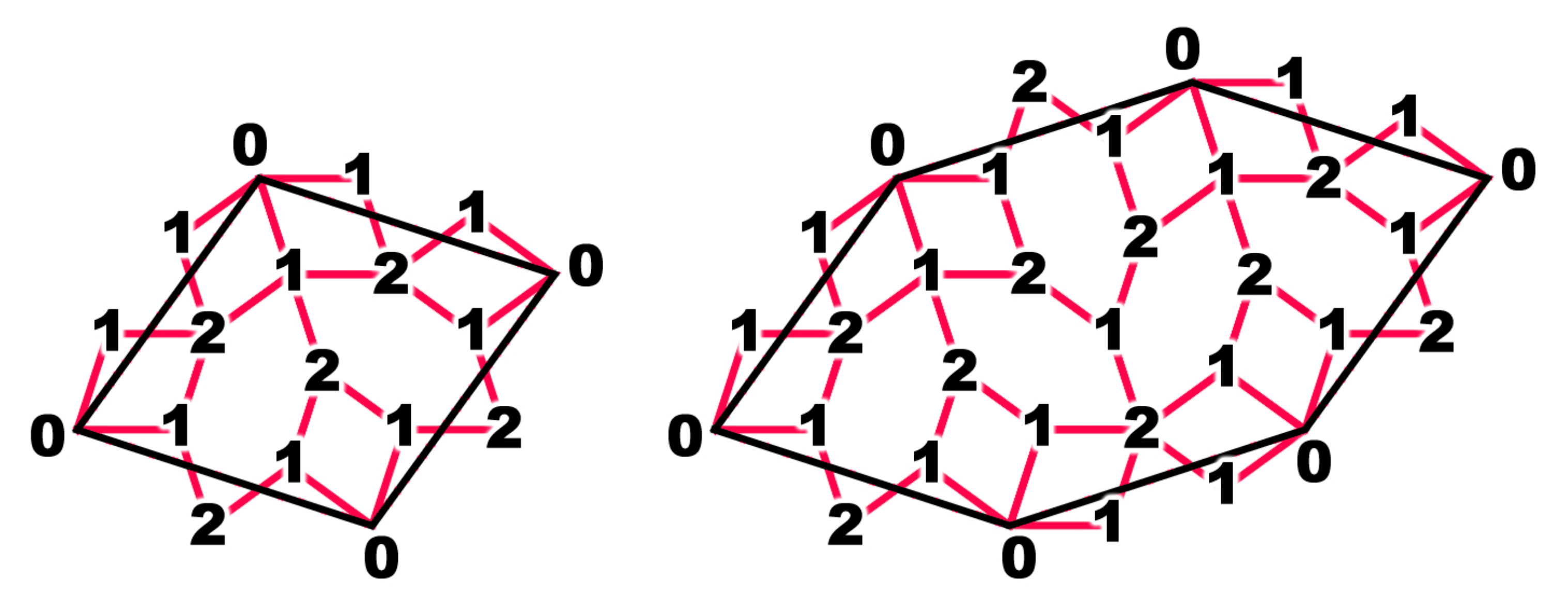}
		\caption{\label{fig:zperp} Vertices of the inflated tiling and their $z_\perp$ coordinates. The tiling can only contain nodes whose projections in $E_\perp$ lie in the planes $z_\perp = 0, 1, 2$. The planes $z_\perp = 0$ and $z_\perp = 2$ contain centers and ray ends of the stars, respectively, the plane $z_\perp = 1$ contains armpits of the stars and common vertices of three hexes. At each inflation step, all vertices of the previous iteration are projected onto the plane $z_\perp = 0$.}
	\end{center}
\end{figure}

\begin{figure}[h]
	\begin{center}
		\includegraphics[width=7cm]{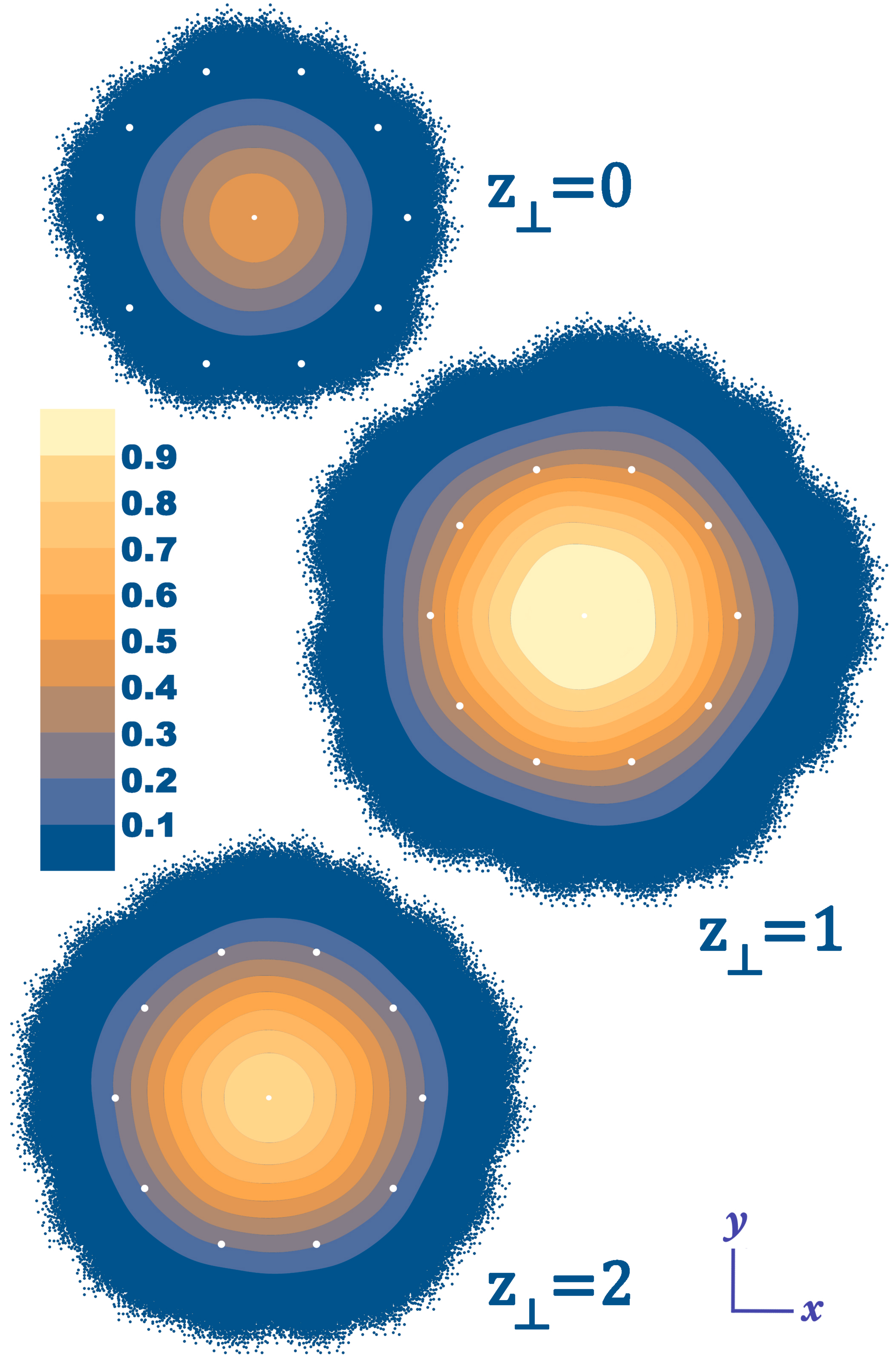}
		\caption{\label{fig:AD} The acceptance domain of the pentagonal tiling from Fig.~\ref{fig:chiral-5fold}, obtained by projecting a region of radius $s_\ell^7 \approx 2615.6$ of the tiling with approximately 21.3 million nodes. Unlike the Henley, Olamy-Kl\'{e}man and Cockayne patterns, this acceptance domain has a non-uniform normalized node density, the levels of which are shown in increments of 0.1. The node density averaged over the polar angle is shown in Fig.~\ref{fig:ADdensity}. The dots show the vertices of decagons of radius $\tau$. Even though the decagons are entirely contained within the boundaries of the acceptance domain, short distances between vertices in physical space do not appear due to the absence of some internal nodes.}
	\end{center}
\end{figure}

\begin{figure}[h]
	\begin{center}
		\includegraphics[width=7cm]{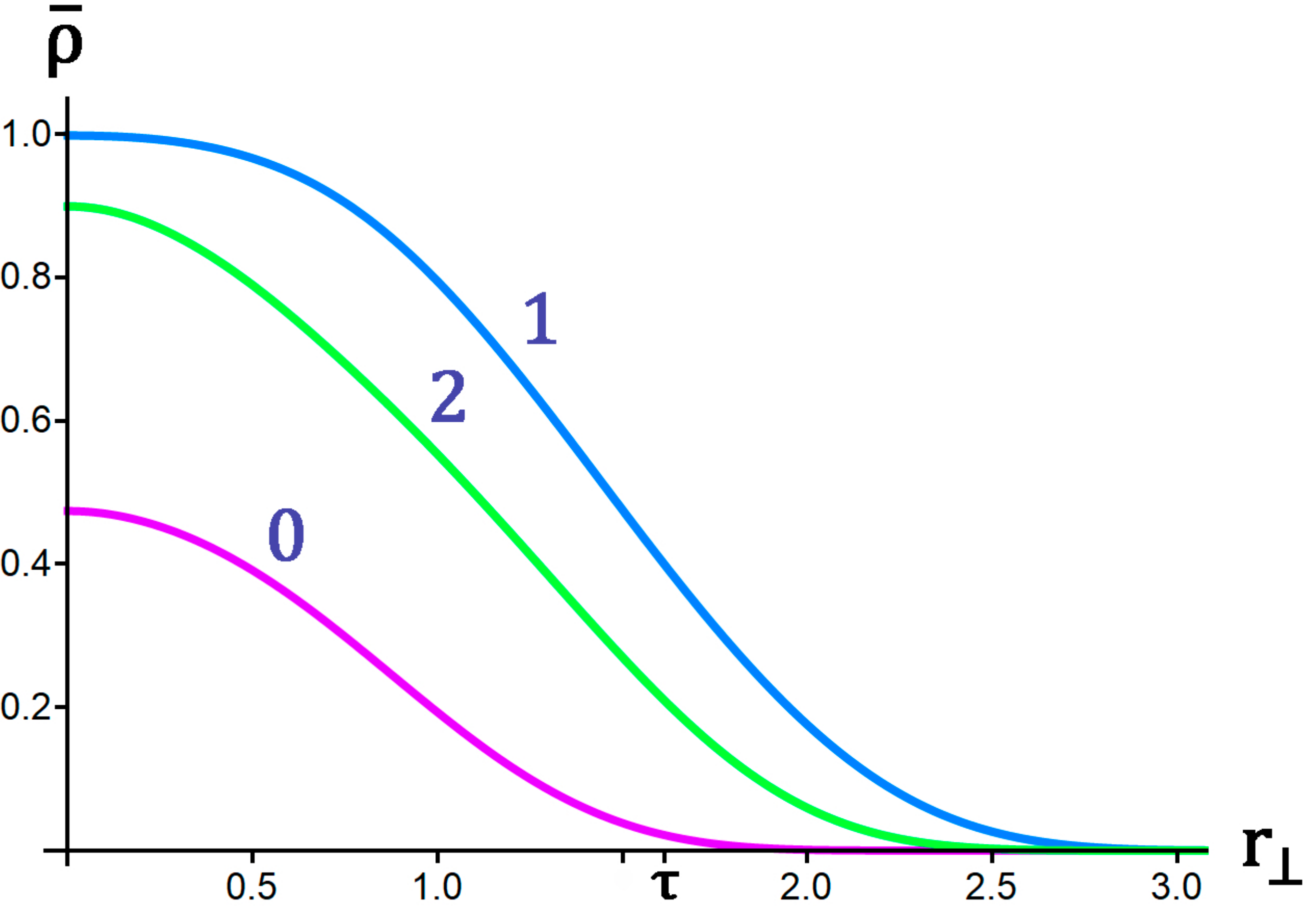}
		\caption{\label{fig:ADdensity} The polar angle-averaged density of the acceptance domain from Fig.~\ref{fig:AD}, calculated for the planes $z_\perp = 0, 1, 2$. Density $\rho = 1$ corresponds to the complete filling. The integral area of the acceptance domain in the plane $z_\perp = i$  can be calculated as $S_i = 2 \pi \int_0^\infty \bar{\rho}_i (r_\perp) r_\perp dr_\perp$. Eqs.~(\ref{eq:S-relative}) and (\ref{eq:Sigma}) are fulfilled for $S_0$, $S_1$, $S_2$.}
	\end{center}
\end{figure}

The acceptance domain of the Penrose tiling is defined by the intersection of atomic planes and the projection of a penteract with side of unit length into the space $E_\perp$. In the case of the standard Penrose tiling, the center of the penteract is located exactly in the middle between two planes, which corresponds to parameter $\varpi=\frac12$ \cite{Senechal1995}. By varying parameter $\varpi$, one can obtain generalized Penrose tilings that are not locally isomorphic to each other. For example, Figure~\ref{fig:penteract} shows the acceptance domains for $\varpi=\frac12$ and $\varpi=0$. 

For each pentagonal quasiperiodic tiling, the density of vertices in physical space is directly proportional to the total area of the flat shapes that make up its acceptance domain. For generalized Penrose tilings, this area does not depend on the parameter $\varpi$ and is
\begin{equation}
	\label{eq:SigmaP}
	\Sigma_\mathrm{P} = \frac52 (2 + \tau)^\frac32 \approx 17.20 .
\end{equation}
Among all the dense pentagonal packings, the Cockayne packing obviously has the largest acceptance domain,
\begin{equation}
	\label{eq:SigmaC}
	\Sigma_\mathrm{C} = \frac54 \tau (2 + \tau)^\frac32 \approx 13.92 .
\end{equation}
From Eqs.~(\ref{eq:rho-tilings}) and (\ref{eq:SigmaC}), we can determine the specific packing density of disks per unit area of acceptance domain,
\begin{equation}
	\label{eq:rho-ADarea}
\frac\rho\Sigma = \frac{\rho_\mathrm{C}}{\Sigma_\mathrm{C}} = \frac{\pi (2\tau - 1)}{125} \approx 0.05620 .
\end{equation}

Geometric restrictions on the size and shape of the acceptance domain are caused by the need to prevent small distances (less than $a = 1$) between the tiling vertices in physical space. For example, as stated in Ref.~\cite{Olamy1989}, two nodes with the same coordinate $z_\perp$ should not be separated by the 5D vector $(1\bar{1}01\bar{1})$ corresponding to the distance $ \sqrt{2 + \tau} / \tau^2 \approx 0.7265$ in $E_\parallel$. To prevent this short distance from appearing, we can limit the acceptance domain in each of the planes $z_\perp = 0 \ldots 4$ by a decagon of radius $\tau$, the vertices of which correspond to 5D vectors of type $(10010)$. All generalized Penrose tilings, as well as the Olamy-Kl\'{e}man one, satisfy this constraint. However, this size restriction can be overcome by using an acceptance domain with a fractal boundary, such as the pinwheel-like acceptance domain of the Cockayne pattern \cite{Cockayne1994}. Another way to overcome the limitation is to reduce the density of nodes inside the acceptance domain, which is what we will show in this case.

As we just discussed, acceptance domains can vary in shape and total area, and the flat shapes that compose them can be either simple polygons or fractal figures such as pinwheel-like atomic surfaces of the Cockayne tiling \cite{Cockayne1994}. However, as a rule, the shapes have a clear boundary that plays a determinative role, namely, all nodes of 5D space whose projections are inside the shape, and only they belong to the tiling. In the case of the tiling described here, the situation is somewhat different. The boundaries are blurred, and nodes that do not belong to the tiling are dense in the acceptance domain. For example, it is easy to show that in the plane $z_\perp = 0$ there is an everywhere dense set of nodes that cannot belong to the pattern. Indeed, the inflation matrices ${\cal I}_R$ and ${\cal I}_L$ project any node of the 5D lattice into the plane $z_\perp = 0$. At the same time, as can be seen from Fig.~\ref{fig:zperp}, during inflation, new nodes can appear only in the planes $z_\perp = 1$ and $z_\perp = 2$. Thus, the entire acceptance domain of the tiling is contained in three planes $z_\perp = 0, 1, 2$. Consequently, the plane $z_\perp = 0$ cannot contain nodes projected from the planes $z_\perp = 3$ and $z_\perp = 4$. This means that node $\mathbf{n}$ with $z_\perp = \sum_k n_k = 0$ is a subject to the additional condition
\begin{equation}
	\label{eq:z0condition}
	\left( \sum_k (p - k) n_k \right) \% 5 \in \{0,1,2\} ,
\end{equation}
where $p$ is an arbitrary integer. This condition results in a non-uniform node density within the acceptance domain.

Figure~\ref{fig:AD} shows the acceptance domain of the $R$-inflated tiling from Fig.~\ref{fig:chiral-5fold}, obtained by projecting approximately 21.3 million of its vertices into $E_\perp$. It can be seen that the acceptance domain has a non-uniform node density. We can find from Fig.~\ref{fig:zperp} that each inflated star contributes 5, 25 and 15 vertices in the planes $z_\perp = 0, 1, 2$, respectively; each inflated hex gives 2, 11 and 7 vertices. Taking into account that $f_\mathrm{hex} / f_\mathrm{star} = \sqrt{5}$, it is not difficult to calculate the acceptance domain relative areas for these three planes,
\begin{equation}
	\label{eq:S-relative}
	S_0 : S_1 : S_2 = 1 : 2 \tau^2 : 2 \tau .
\end{equation}
Here, the integral area $S_i = \int \rho_i(\mathbf{r}_\perp) d\mathbf{r}_\perp$, $i = 0, 1, 2$, where $\mathbf{r}_\perp$ is in-plane coordinate and $\rho_i$ is the normalized density of vertices ($\rho_i = 1$ corresponds to complete filling). The normalized node density for the planes $z_\perp = 0, 1, 2$ averaged over the polar angle is shown in Fig.~\ref{fig:ADdensity}. The total area of the acceptance domain is the same as that of the Cockayne tiling,
\begin{equation}
	\label{eq:Sigma}
	S_0 + S_1 + S_2 = \Sigma_\mathrm{C} .
\end{equation}

\section{Discussion}
\label{sec:discussion}
HBS tilings are widely used to describe the real structures of decagonal and pentagonal quasicrystals and their periodic approximants \cite{Cockayne1998,Cockayne2000,Steurer2001,Henley2002,Mihalkovic2002,Reichert2003,Gummelt2004,Widom2004,Mihalkovic2004,He2017}. Along with the deterministic Henley, Olamy-Kl\'{e}man, and Cockayne patterns discussed in Sec.~\ref{sec:history}, an infinite number of random coverings can be constructed from these three types of tiles \cite{Cockayne1998,Reichert2003,Gummelt2004}. The entropy induced by random defects of the atomic structure, such as phasonic jumps, plays a significant role, making the quasiperiodic order more favorable than the periodic one \cite{Fayen2024}. Monte Carlo simulations of quasicrystal growth also confirm their structural imperfection. In particular, grown {\it in silico} icosahedral quasicrystals have a nearly spherical acceptance domain with a loose structure \cite{Dmitrienko1995PRL,Dmitrienko1995JETPL,Dmitrienko1999}.

\begin{figure}[h]
	\begin{center}
		\includegraphics[width=5.25cm]{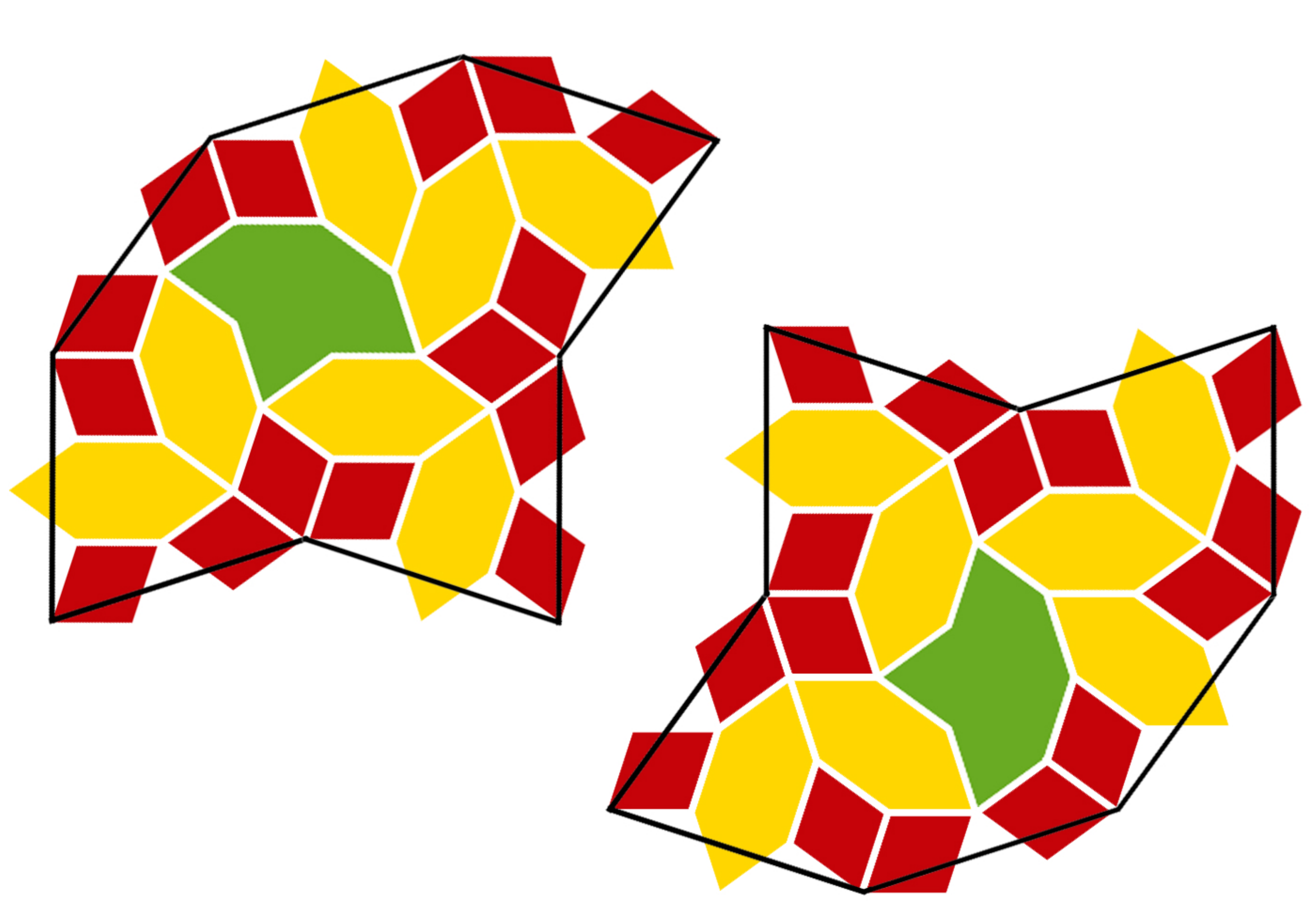}
		\caption{\label{fig:ivy-leaf} Inflation rules for two of the ten possible orientations of ivy leaves, consistent with the matching rules from Fig.~\ref{fig:inflation}. During the inflation process, each ivy leaf produces exactly one leaf of the next generation, so their number is preserved, and their relative frequency of appearance in the tiling decreases by about $s_\ell^2$ times at each step of inflation.}
	\end{center}
\end{figure}

\begin{figure}[h]
	\begin{center}
		\includegraphics[width=7cm]{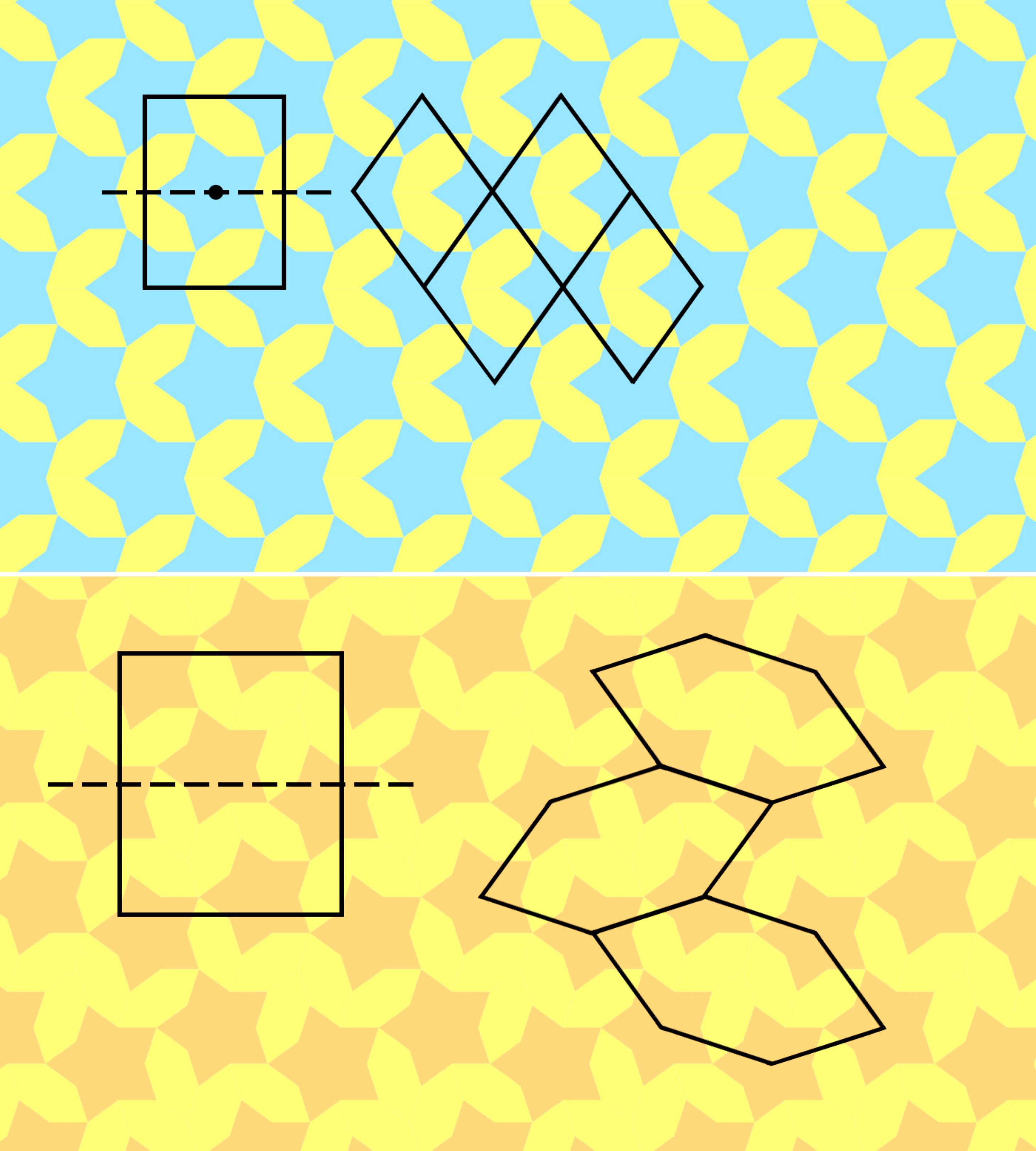}
		\caption{\label{fig:approximants} Periodic approximants of pentagonal HS tilings obtained by applying inflation once to tilings of rhombuses (top) and hexagons (bottom). The symmetry of the patterns is described by the plane groups $cm$ and $pg$, respectively. The unit cells of the periodic structures are shown.}
	\end{center}
\end{figure}

In this paper, inflation rules were proposed for a chiral pentagonal tiling of starfish and hexes (HS-tiling). Although at each step the inflation rules can be chosen in two ways: $R$ or $L$, this does not make a significant contribution to disorder. Thus, the structure is almost deterministic. However, this conclusion is valid only for HS-tilings made using an infinite number of inflation steps, that is, those that are described by an infinite sequence of symbols $R$ and $L$, e.g. $RRLRLLR\ldots$ Note that the inflation rules in Fig.~\ref{fig:inflation} are actually defined for hexes and rhombi with an acute angle of $2\pi/5$. Therefore, any tilings composed of these two shapes may be subject to inflation, including high entropy Cockayne patterns \cite{Cockayne1994}. In symbolic form, the tilings obtained in this way can be written as a finite sequence, for example  $RRL\ast$, where the asterisk denotes the original tiling of rhombuses and hexes, which itself cannot be obtained from another tiling using the inflation rules from Fig.~\ref{fig:inflation}.

Ivy leaves (boats) may also be included in consideration as of HS pattern defects. Figure~\ref{fig:ivy-leaf} shows the inflation rules for ivy leaves, consistent with the matching rules from Fig.~\ref{fig:inflation}. With inflation, the number of ivy leaves does not change, but their frequency of appearance in the tiling decreases promptly compared to starfish and hexes.

The method also makes it possible to obtain structures of periodic approximants of pentagonal quasicrystals, starting from periodic tilings of rhombi and hexes. Figure~\ref{fig:approximants} shows examples of two periodic patterns of starfish and hexes obtained by applying inflation once.

\section*{Acknowledgements}
The author is grateful to V.~E.~Dmitrienko and M.~V.~Gorkunov for fruitful discussions that led to the idea for this paper. The work was carried out within the state assignment of NRC ``Kurchatov Institute''.


\end{document}